\documentclass[12pt]{article}
\usepackage{graphicx}
\input epsf
\begin{document}
\font\bss=cmr12 scaled\magstep 0
\title{The Generalized Fubini instanton}
\author{A.A. Yurova*, A.V. Yurov** \\
\small *Department of  Higher Mathematics\\
\small Kaliningrad State Techical University\\
\small 236000, Soviet Avenue, 1, Kalinigrad, Russia, yurov@freemail.ru\\
\small **Department of Theoretical Physics\\
\small Russian State University of I. Kant\\
\small 236041, Aleksandra Nevskogo street, 14, Kalinigrad, Russia,
artyom\_yurov@mail.ru }
\date {}
\maketitle
\begin{abstract}
We show that $(1+2)$ nonlinear Klein-Gordon equation with negative
coupling admits an exact solution which appears to be the linear
superposition of the plane wave and the nonsingular rational soliton.
We show that the same approach allows to construct the
solution of similar properties for the Euclidean $\phi^4$ model with broken
symmetry. Interestingly, this regular solution will be of instanton type
only in the $D\le 5$ Euclidean space.
\end{abstract}
\thispagestyle{empty}
\medskip
\section{Introduction}

The nonlinear Klein-Gordon equation ($\phi^4$- or
$|\phi|^4$-models) belongs to the class of those equations that frequently arises in seemingly
 different physical applications: see, for example, \cite{Vilenkin} (cosmic strings),
\cite{Yang} (field theory), \cite{Manton} (condensed matter
physics, e.g. liquid crystals).

Unfortunately, this important equation is not an integrable one. In the
particular case of $(1+1)$ one can still obtain stationary solutions (for example,
kinks for the $\phi^4 $-model), but in the case of general position
$(1+D)$ there are no general methods allowing one to construct
the exact solutions. There are some exceptions, however: most notably in the cases when
the required solutions have sufficient number of symmetries. Such symmetries often allow
to reduce the problem to the one-dimensional one and to successfully integrate the equation.
For example, in the massless theory with the potential
\begin{equation}
V(\phi)=-\frac{\lambda\phi^4}{4}, \label{1.1}
\end{equation}
and upon introduction of the special boundary conditions the
Euclidean solution with $O(4)$ symmetry can be obtained. This
solution, the so-called Fubini instanton is one-parametric,
nonsingular, and has the form $\phi(r)$ ($r=\sqrt{x_i^2}$, the
$x_i$ are the Euclidean coordinates) and finite Euclidean action
\cite{Fubini}. A somewhat surprising fact is that for any
non-vanishing $m$ and the potential
\begin{equation}
V(\phi)=\frac{m^2\phi^2}{2}-\frac{\lambda\phi^4}{4}, \label{poten}
\end{equation}
there are no instanton solutions in the theory. This is the good example of the nontrivial properties of solutions for nonlinear multidimensional equations like the mentioned $\phi^4$-model.

In this article we are going to construct the generalization of the
Fubini instanton (FI), i.e. we are going to find out the solution of the
Euclidian equation of arbitrary dimensions and with the non-vanishing
$m$. As we shall see, this task is accomplishable for the generalized potentials of the kind:
$$
V(\phi)=V_0+\alpha\phi+\frac{m^2}{2}\phi^2+\frac{\beta}{3}\phi^3-\frac{\lambda}{4}\phi^4.
$$
To be instanton this solution should be generated by the very special boundary
conditions and must result in finite Euclidean action. As we shall see this will be the case only for the
Euclidean spaces with $D\le 5$.

In order to avoid any possible confusion, let us make one little remark: everywhere throughout
this article the $\lambda$ (coupling constant) is assumed to be a positive quantity ($\lambda>0$).
Since the models we are going to discuss will (with just a couple of exceptions) all have the negative coupling constants,
our restriction will just mean the presence of the negative sign in front of the
corresponding terms (see, for example, equations (\ref{1.1}) and (\ref{poten})).

The plan of the article is as follows: first of all, we are going to consider the $(1+D)$  $|\phi|^4$-model in
the Minkowski space (summation is implied over repeating contravariant and covariant indices):
\begin{equation}
\partial_{\mu}\partial^{\mu}\phi+m^2\phi-\lambda|\phi|^2\phi=0,
\label{equation}
\end{equation}
with the metric
$$
g_{\mu\nu}={\rm diag}(+1,-1,-1,...,-1).
$$
As we shall see, in the case $D=2$  the equation
(\ref{equation}) admits the nonsingular solutions $\phi(x^{\mu})$,
$\mu=0,\,1,\,2$ such that
$$
|\phi(x^{\mu})|\to B={\rm const}\qquad {\rm at}\qquad
x^2+y^2\to\infty.
$$
We assume that similar solutions can be obtained for the case $D>2$ also, but this is still a hypothesis.

Of course, at first sight the models with negative coupling seems
to bear little to no connection with the reality. In fact, since
the potential is negative, the action is not bounded from below
and therefore, it might not be stable quantum mechanically.
However, it appears that models with the negative potentials can
be very important in such areas as cosmology and strings theory
\cite{Linde}. Thus, the problem to find interesting and nontrivial
exact solutions of such models should be regarded as an actual
one.

We are discussing such solutions because they somehow hints the
correct anzats for the model we are interested in with the real
scalar filed in the euclidean space. We will show (in Sec.4) that
by the usage of the very simple ideas it is possible to find out
the generalization of the FI - which is specifically the main aim
of this article. This solution is valid for the Euclidian space of
arbitrary number of dimensions and arbitrary sign in front of
$\lambda$. However, to obtain a regular (i.e. without
any peculiarities) solution one need to choose the negative sign.

The most amazing fact here is that of all the possible dimensions
of the model, the dimension $D=4$ stands out. In particular, case
$D=4$ is the only one that requires a special treatment, which
treatment we are going to discuss in section 5.

\section{Equations}
In \cite{Yurov} the novel exact solution of the Davey-Stewartson
II (DS-II) equations describing the soliton on the plane wave
background is obtained. This solution has the form
\begin{equation}
\phi(x,y,t)=B{\rm
e}^{is(x,y,t)}\left(-1+\frac{P_1(x,y,t)}{P_2(x,y,t)}\right),
\label{anzatz}
\end{equation}
where $s(x,y,t)$ and $P_1(x,y,t)$ are linear functions whereas
$P_2(x,y,t)$ is polynomial of the second order such that
$P_2(x,y,t)>0$ for any values of $x$, $y$, $t$.

The aim of this work is to show that the nonintegrable (1+2)
$|\phi|^4$ with negative coupling admits the similar solution (see
(\ref{anzatz})) with
\begin{equation}
\begin{array}{l}
P_1=a_{\mu}x^{\mu}+a,\qquad P_2=\eta_{\mu\nu}x^{\mu}x^{\nu}+b_{\mu}x^{\mu}+A^2,\\
\\
s=s_{\mu}x^{\mu},
\end{array}
\label{forma}
\end{equation}
where
$$
\begin{array}{l}
\eta_{\mu\nu}=\eta_{\nu\mu},\qquad
\eta_{\mu\nu}=\left(\eta_{\mu\nu}\right)^*,\\
\\
(A^2)^*=A^2, \qquad B^*=B,\qquad (s_{\mu})^*=s_{\mu}.
\end{array}
$$
The solution (\ref{anzatz}) will be nonsingular if
\begin{equation}
\begin{array}{c}
\eta_{11}>0,\qquad
\left|\begin{array}{cc} \eta_{11}&\eta_{12}\\
\eta_{12}&\eta_{22}
\end{array}\right|>0,\\
\\
\left|\begin{array}{ccc} \eta_{00}&\eta_{01}&\eta_{02}\\
\eta_{01}&\eta_{11}&\eta_{12}\\
\eta_{02}&\eta_{12}&\eta_{22}
\end{array}\right|>0,\qquad
\left|\begin{array}{cccc} 0&b_{0}&b_{1}&b_{2}\\
b_{0}&\eta_{00}&\eta_{01}&\eta_{02}\\
b_{1}&\eta_{01}&\eta_{11}&\eta_{12}\\
b_{2}&\eta_{02}&\eta_{12}&\eta_{22}
\end{array}\right|>0.
\end{array}
\label{nerav}
\end{equation}
Substituting (\ref{forma}) into (\ref{equation}) one gets:
\begin{equation}
\begin{array}{l}
iP_2\left[P_2\left(P_1-P_2\right)\partial_{\mu}\partial^{\mu}s+2J^{\mu}\partial_{\mu}s\right]+
P_2\partial_{\mu}J^{\mu}-2J^{\mu}\partial_{\mu}P_2+\\
\\
+\left(P_1-P_2\right)\left[-\lambda
B^2(P_1-P_2)(P_2-P^*_1)+(m^2-\partial_{\mu}s\partial^{\mu}s)P_2^2\right]=0,
\end{array}
\label{uh}
\end{equation}
where
$$
J^{\mu}=P_2\partial^{\mu}P_1-P_1\partial^{\mu}P_2.
$$
 Using (\ref{uh}) for the particular case $b_{\mu}=0$ one
will end up with the following system
\begin{equation}
\begin{array}{l}
s_{\mu}s^{\mu}+\lambda B^2-m^2=0,\\
\\
(2is_{\mu}a^{\mu}-\lambda B^2(a+a^*))A^4-a(-\lambda
B^2(a+2a^*)+2\eta^{\mu}_{\mu})A^2-\lambda B^2|a|^2a=0,\\
\\
a_{\mu}+a^*_{\mu}=0,\\
\\
\left(2is_{\mu}a^{\mu}-\lambda
B^2(a+a^*)\right)\eta_{\alpha\beta}-4is^{\mu}\eta_{\mu\alpha}a_{\beta}-\lambda
B^2a_{\alpha}a_{\beta}=0,\\
\\
2\eta_{\alpha\beta}\left[\left(-\lambda B^2
a^*+\eta^{\mu}_{\mu}\right)a_{\rho}+2\left(a_{\mu}+ias_{\mu}\right)\eta^{\mu}_{\rho}\right]+
a_{\rho}\left[-\lambda B^2
a_{\alpha}a_{\beta}-8\eta_{\mu\alpha}\eta^{\mu}_{\beta}\right]=0,\\
\\
\left[-\left(a+2a^*-2A^2(a+a^*)\right)\lambda
B^2+8iaA^2\eta^{\mu}_{\mu}s^{\nu}a_{\nu}\right]\eta_{\alpha\beta}+\lambda
B^2\left(A^2+a^*-2a\right)a_{\alpha}a_{\beta}-\\
-8a\eta_{\mu\alpha}\eta^{\mu}_{\beta}+4iA^2s^{\mu}\eta_{\mu\alpha}a_{\beta}=0,\\
\\
\left[-\lambda
B^2\left(2a^*A^2-2|a|^2+a^2\right)+2\eta^{\mu}_{\mu}A^2\right]a_{\alpha}+4A^2\left(a^{\mu}+ias^{\mu}\right)\eta_{\mu\alpha}=0.
\end{array}
\label{sys}
\end{equation}
Thus for the case $(1+D)$ one should solve the system of
$$
N(D)=\frac{(D+2)(2D^2+5D+7)}{2}
$$
algebraic equations with two additional conditions:
\newline
(i) the inequalities (\ref{nerav}) must hold and
\newline
(ii) level lines of the function $P_2$ must be the closed curves.

\section{$D=2$ Solutions}

In the case $D=2$: $N(D)=50$. Assuming $\eta_0$, $\alpha$, $\beta$,
$\gamma$, $\rho$ and $b$ to be the arbitrary real parameters, define
three Lorentzian vectors
$$
\xi_{\mu}=(\xi_0,\xi_1,\xi_2),\qquad
\eta_{\mu}=(\eta_0,\eta_1,\eta_2),\qquad
\theta_{\mu}=(\theta_0,\theta_1,\theta_2),
$$
such that
\begin{equation}
\begin{array}{l}
\displaystyle{ \eta_1=\eta_0\cos\,\alpha,\qquad
\eta_2=\eta_0\sin\,\alpha,\qquad
\sigma=\frac{\rho^2\sin[2(\beta-\alpha)]}{4\eta_0\sin(\alpha-\gamma)},}\\
\\
\xi_1=\rho\cos\,\beta,\qquad \xi_2=\rho\sin\,\beta,\qquad
\xi_0=\rho\cos(\beta-\alpha),\\
\\
\displaystyle{ \theta_1=\sigma\cos\,\gamma,\qquad
\theta_2=\sigma\sin\,\gamma,\qquad
\theta_0=\frac{\sigma\cos(\beta-\gamma)}{\cos(\beta-\alpha)}, }
\end{array}
\label{vectors}
\end{equation}
and besides
\begin{equation}
\lambda=2\rho^2\sin^2(\alpha-\beta),\qquad
m^2=\frac{\sigma^2\sin(2\beta-\gamma-\alpha)\sin(\gamma-\alpha)}{\cos^2(\beta-\alpha)}.
\label{couplings}
\end{equation}
 Then the  solution of the system (\ref{sys}) such the
(\ref{nerav}) and two conditions (i) and (ii) holds has the
form:
\begin{equation}
\begin{array}{l}
\eta_{\mu\nu}=\xi_{\mu}\xi_{\nu}-2b\xi_{\mu}\eta_{\nu}+4\left(b^2+B^2\right)
\eta_{\mu}\eta_{\nu},\qquad
s_{\mu}=\left(2B^2-b^2\right)\eta_{\mu}+b\xi_{\mu}+\theta_{\mu},\\
\\
\displaystyle{ a_{\mu}=4i\eta_{\mu},\qquad a=\frac{1}{B^2},\qquad
A=\pm\frac{1}{2B}.}
\end{array}
\label{solution1}
\end{equation}
This would be the case if $b_{\mu}=0$. If $b_{\mu}\ne 0$ then one have
a slightly more complicated system them (\ref{sys}). The particular solution
can be obtained, though:
\begin{equation}
\begin{array}{l}
\displaystyle{
\eta_{\mu\nu}=4B^2\left(\xi_{\mu}\xi_{\nu}-2b(\xi_{\mu}\eta_{\nu}+\xi_{\nu}\eta_{\mu})+
4\left(b^2+B^2\right)\eta_{\mu}\eta_{\nu}\right),}\\
\\
\displaystyle{b_{\mu}=8B^2\left(2(\kappa\psi B+\chi
b)\eta_{\mu}-\chi\xi_{\mu}\right),\qquad
a_{\mu}=16iB^2\eta_{\mu},}
\\
\\
\displaystyle{ a=4\left(1+\frac{\kappa
B}{|c_1|^2}(c_2c^*_1-c^*_2c_1)\right),\qquad A^2=4B^2(\chi^2+\psi^2),}\\
\\
\displaystyle{
\chi=\frac{\kappa|c_1|^2-B(c_1c^*_2+c^*_1c_2)}{2B|c_1|^2},\qquad
\psi=\frac{i(c_1c^*_2-c^*_1c_2)}{2|c_1|^2},}
\end{array}
\label{solution2}
\end{equation}
where $c_{1,2}$ $\kappa=\pm 1$ are the arbitrary complex constants
and $s_{\mu}$ is similar to the previous one (see
(\ref{solution1})).

In order to show that (\ref{nerav}) and two conditions (i) and
(ii) hold it is sufficient to show the following:
$$
\eta_{\mu\nu}x^{\mu}x^{\nu}+b_{\mu}x^{\mu}+A^2=1+4B^2\left[\left(-\Lambda_{\mu}x^{\mu}-
\chi\right)^2+\left(\L_{\mu}x^{\mu}+ \psi\right)^2\right],
$$
where
$$
-\Lambda_{\mu}=\xi_{\mu}-2b\eta_{\mu},\qquad L_{\mu}=2\kappa
B\eta_{\mu}.
$$

{\bf Remark 1}. In articles \cite{Taj1}, \cite{Taj2} we has
presented the method of construction of the exact solutions of
equation (\ref{equation}), based on the usage of the self-similar
change of variables, effectively reducing (\ref{equation}) to the
nonlinear Schr\"odinger equation. It is possible to show that each
solution of NLS corresponds to the solution of (\ref{equation})
(the contrary statement is certainly false). In particular, using
the multi-soliton solutions of the NLS one can find out the
corresponding ``multi-soliton'' solutions of (\ref{equation}).
Geometrically, such solitons are similar to those of the
Kadomtsev-Petviashvili equation with the streamlines being just a
familiy of a parallel straight lines. Unlike to the solutions,
constructed in the works \cite{Taj1, Taj2}, the (\ref{anzatz})
appears to be the linear superposition of the plane wave and the
rational soliton with level lines having the form of ellipses on
the plane $xy$.

\section{Generalized Fubini instantons in the 5D model with broken symmetry}

We have already noted that in the $\phi^4$ model with potential
(\ref{poten}) for any nonvanishing $m$ there exist no instanton solutions. On the other hand, we have shown that the complex model (\ref{equation}) does have the nonsingular solutions with the following behaviour:
$$
|\phi(x^{\mu})|\to {\rm const}\qquad {\rm at}\qquad
x^2+y^2\to\infty.
$$
Taking in regard the similarity between (\ref{poten}) and
(\ref{equation}) models (both of them has the nonlinearity of the
fourth order) it is natural to search for the regular solutions of
the real-valued model $\phi^4$ with the similar behaviour. In this
section we will show that such solutions indeed exist and that
their Euclidean action is bound to be finite, i.e. such solutions
are in fact instantons. However, this would be true for the 5D
space-time only.

The D=4 Fabini instantons has the form
\begin{equation}
\phi(r)=2\sqrt{\frac{2}{\lambda}}\frac{\rho}{r^2+\rho^2},
\label{Fabini}
\end{equation}
where $\rho$ is arbitrary. The corresponding Euclidean action is
$$
S^{(4)}_E=\frac{8\pi^2}{3\lambda}.
$$
Such solutions are interesting in studies of the the tunneling
trajectories. All we need now are two boundary conditions:
\newline
(i) $\phi(r=\infty)=\phi_0$ and
\newline
(ii) $d\phi/dr|_{r=0}=0$.

Instantons (\ref{Fabini}) are the solutions whose minimal action
posses the $O(4)$ symmetry of Euclidean space and which are
created by the boundary conditions (i) and (ii).

In the previous sections we have shown that (\ref{equation})
admits a certain type of solitons with the boundary condition akin
to (i). Although the calculations so far has been done only for
the (1+2) case, it seems natural to expect the similar solutions
in the higher dimensions. Let's look for the generalization of
(\ref{Fabini}) having the following form:
\begin{equation}
\phi(x^i)=\phi_0+\frac{a}{a_{ij}x^{i}x^j+b_ix^i+\rho^2},
\label{anzatz}
\end{equation}
where the Euclidean variables has been renumbered with the aid of
the Latin indexes ($i,j=1...D$). Unfortunately, the direct check
here would be extremely cumbersome, so we'll restrict ourselves to
the case of the $O(D)$ spherical symmetry. In this particular case
the Euclidean equations will have the form:
\begin{equation}
\frac{d^2\phi}{dr^2}+\frac{D-1}{r}\frac{d\phi}{dr}=\frac{dV(\phi)}{d\phi}.
\label{uur}
\end{equation}
Since the number of the free parameters will get too restricted,
we'll add such parameters by introduction of the broken symmetry
into the model:
\begin{equation}
V(\phi)=V_0+\alpha\phi+\frac{m^2\phi^2}{2}+\frac{\beta\phi^3}{3}-\frac{\lambda\phi^4}{4}.
\label{potencial}
\end{equation}

Using the ansatz (\ref{anzatz}) by complete analogy with the
(\ref{equation}) we'll end up with the required
solution.~\footnote{We omit most of the calculations made in this
sections due to the fact that they are essentially the same as the
previously performed ones}:
\begin{equation}
\phi(r)=\phi_0+\frac{4\phi_0(D-4)}{\mu\left(r^2+\rho^2\right)},
\label{inst}
\end{equation}
where
$$
\rho^2=\frac{2\lambda\phi_0^2(D-4)^2}{\mu^2},\qquad
\alpha=-\frac{\phi_0(\lambda \phi_0^2+m^2)}{2},
$$
$$
\beta=\frac{3\lambda\phi_0^2-m^2}{2\phi_0},\qquad
\mu=3\lambda\phi_0^2+m^2,
$$
where $\phi_0$ is arbitrary. One can see that both (i) and (ii)
are hold. The calculation results in
$$
\lim_{r\to\infty}\left(\frac{1}{2}\left(\frac{d\phi}{dr}\right)^2+V(\phi)\right)=V_0-\frac{\phi_0^2(3\lambda\phi_0^2+2m^2)}{12},
$$
so in order to obtain the finite Euclidean action one should take
$$
V_0=\frac{\phi_0^2(3\lambda\phi_0^2+2m^2)}{12}.
$$
Then the Lagrangian will have the following asymptotic:
$$
\frac{1}{2}\left(\frac{d\phi}{dr}\right)^2+V(\phi)\to r^{-6},
$$
when $r\to\infty$. In order to calculate the action one has to
multiply the Lagrangian by $r^{D-1}$. It is easy to see that the
Euclidean action would indeed be finite only in the case $D\le 5$.
It seems that $D<4$ cases are physically meaningless and provide
little interest for the quantum cosmology (see, however, the note
2). Case $D=4$ requires the specific treatment and will be
considered in the next section. Hence, the only case left for us
to check out is the one with $D=5$, whose Euclidean action has the
form:
$$
S^{(5)}_E=\frac{8\pi^2}{3}\int_{0}^{\infty}
r^4\left(\frac{1}{2}\left(\frac{d\phi}{dr}\right)^2+V(\phi)\right)dr=\frac{8\pi^3\phi_0\sqrt{2}}
{3\sqrt{\lambda}(3\lambda\phi_0^2+m^2)}.
$$
and the probability of the bubble formation can be estimated as
$$
P\sim\exp\left(-\frac{8\pi^3\phi_0\sqrt{2}}
{3\sqrt{\lambda}(3\lambda\phi_0^2+m^2)} \right).
$$
\newline
{\bf Remark 2}. Multiplying the (\ref{inst}) by the oscillating
exponents and making such choice of the parameters as to set the
cubic term ($\beta=0$) to zero, we'll end up with the solution of
the non-homogeneous multidimensional NLS:
\begin{equation}
i\Psi_t+\Delta^{(D)}\Psi+\left(\kappa-3\lambda
c^2\right)\Psi+\lambda|\Psi|^2\Psi=-2\lambda c^3 {\rm e}^{i\kappa
t}, \label{NLS}
\end{equation}
where $\Delta^{(D)}$ is a $D$-dimensional Laplacian and $\kappa$,
$c$ are arbitrary constants. An exact solution of such equation
will be of a form:
\begin{equation}
\displaystyle{ \Psi={\rm e}^{i\kappa
t}\left(c+\frac{2(D-4)}{3c\lambda\left(\sum_{k=1}^D
x_k^2+\frac{(D-4)^2}{18\lambda c^2}\right)}\right).}
 \label{NLS-sol}
\end{equation}
By a particular choice $\kappa=3\lambda c^2$ it is possible to
reduce the left hand side of (\ref{NLS}) to the traditional NLS
form. This solution works for both focusing and defocusing NLS,
but for the latter the equation will have a singularity. It is
well known that in case of $D>1$ and under the assumption of the
non-vanishing right hand side the equation (\ref{NLS}) becomes an
non-integrable one, posing quite a challenge for those who dares
to find out its exact solutions. On the other hand, NLS is the one
of the most commonly used equations in physics. So we hope that
the existence of the exact solution (\ref{NLS-sol}) for equation
(\ref{NLS}) would be useful in different physical applications.

\section{Case $D=4$}
The Bogolyubov transformation $\phi\to\phi+c$ is the one commonly
applied for the models with the spontaneously broken symmetry,
i.e. for the case when either Lagrangian or Hamiltonian allow for
some kind of symmetry, while the ground state is not invariant
with respect to the corresponding symmetric transformation.
However, the Bogolyubov transformation can also be applied in a
slightly more general context. Lets take for example the solution
(\ref{inst}). Since as $r\to\infty$, $\phi\to\phi_0$, it is quite
natural to try a new field variable $\Phi=\phi-\phi_0$, which goes
to zero as $r\to\infty$.

Lets consider the following simple potential
\begin{equation}
V(\Phi)=\frac{\lambda\Phi^4}{4}+V_0', \label{5.1}
\end{equation}
where $V_0'$=const (we remind our readers, that $\lambda>0$, i.e.
the model under description is the one with the positive coupling
constant). Introducing the new variable $\phi$, we end up with
\begin{equation}
V(\phi)=\frac{\lambda\phi^4}{4}-c\phi_0\phi^3+\frac{3\lambda\phi_0^2}{2}\phi^2-\lambda\phi_0^3\Phi+V_0''.
\label{5.2}
\end{equation}

Comparing (\ref{5.2}) and (\ref{potencial}) (with regard to the
$\lambda\to -\lambda$ substitution), we deduce the expression for
the mass
\begin{equation}
m=\sqrt{3\lambda}\phi_0. \label{5.3}
\end{equation}
The direct substitution of (\ref{5.3}) into (\ref{inst}) will
obviously result in singularity. In order to avoid this, let us
point out that for $D=4$ the expression contains the
indeterminacy, whose elimination might in principle give us the
reasonable result. Hence, the ``symmetry breaking''
procedure~\footnote{This can be understood literally: while
initial model (\ref{5.1}) has had the discrete reflection symmetry
$\Phi\to-\Phi$, the (\ref{5.2}) does not. However, it is more
convenient to refer to this as to the ``hidden'', not ``broken''
symmetry: after all, the later term traditionally has a precise
and strict definition - see the beginning of this chapter.} is
possible for the $D=4$ only.

Now introduce the following handy substitution:
\begin{equation}
m^2-3\lambda c\phi_0^2=\epsilon(D-4) \label{5.4}
\end{equation}
and take the limit $\epsilon\to 0$. Substituting (\ref{5.4}) into
(\ref{inst}) will result in:
\begin{equation}
\Phi=\phi_0=\frac{4\epsilon\phi_0}{\epsilon^2r^2-2\lambda
\phi_0^2}. \label{5.5}
\end{equation}
Keeping $\epsilon$ as a finite quantity, it is possible to
substitute (\ref{5.4}) directly into (\ref{uur}) with the
potential (\ref{5.2}) and to convince oneself that for $D=4$ the
(\ref{5.4}) is the solution indeed. Let us emphasize that this
solution corresponds to the case of nonzero mass, but is singular
and, hence, leads to certain problems with calculations on the
Euclidean action and the probability of the bubble formation.
However, there are still some ways to bypass those difficulties.
First, if we'll assume in (\ref{5.2}) $V_0''=\lambda\phi_0^4/4$ it
will give us
$$
\lim_{r\to\infty}\left(\frac{1}{2}\left(\frac{d\phi}{dr}\right)^2+V(\phi)\right)=0.
$$
Second, calculating the lagrangian
$$
L=\frac{32\phi_0^2\epsilon^2\left(r^2\epsilon^2+2\lambda
c^2\right)}{\left(r^2\epsilon^2-2\lambda c^2\right)},
$$
and substituting it into the expression for the Euclidean action
\begin{equation}
S_E^{(4)}=2\pi^2\int_0^{\infty} r^3Ldr. \label{S-4}
\end{equation}
will end up in expression, having the inherent singularity (pole
of the fourth order). However, the troublesome term can be
effectively separated from other ones by representing (\ref{S-4})
as:
\begin{equation}
S_E^{(4)}=-\frac{8\pi^2}{3\lambda}+S_{\infty}^{(4)},
\label{inftyy}
\end{equation}
where
$$
S_{\infty}^{(4)}=\frac{8\pi^2}{3\lambda}\lim_{\epsilon\to
0}\frac{1}{\epsilon^3}.
$$
Note that the probability of the bubble formation is defined by
\begin{equation}
P=\exp\left(\frac{8\pi^2}{3\lambda}-S_{\infty}^{(4)}\right)=
N_{\infty}\exp\left(\frac{8\pi^2}{3\lambda}\right). \label{555}
\end{equation}
Note also, that the emergence of the ``strange'' multiplicative
constants, such as $N_{\infty}$ and alike during the calculations
of amplitudes by the functional integral method is quite common
and should not be highly surprising \cite{Ramon}. Following the
generally accepted (although not strictly proven) methods, one can
simply cancel the $N_{\infty}$ quantity in the $P$. As a result,
we'll get the expression
$$
P\sim\exp\left(\frac{8\pi^2}{3\lambda}\right),
$$
which is similar to the expression, obtained for the normal Fubini
instanton \cite{Linde-1} up to the sign .

The problems, described above are the inherent ones for the models with the positive coupling constant. In case of the negative coupling constant $\lambda\to-\lambda$ the four-dimensional action has the normal form
\begin{equation}
S_E^{(4)}=\frac{8\pi^2}{3\lambda}, \label{inftyy-1}
\end{equation}
and provides exactly the same probabilities as the massless model with the Fubini
instanton.

\section{Conclusion}
 There are three open questions which are
interesting problems for the future investigation.

The first question is the generalization of the method to the case
$D>2$. At a moment there are no obvious reasons to prevent us from
using our approach in the multi-dimensional case. The only price
we would have to pay working there would be the rapidly growing
amount of calculations. In fact, the number of algebraic equations
$N(D)$ which should be solved grows as $D^3$: $N(3)=100$,
$N(4)=177$ and $N(10)=1542$. Our point about this is following:
even if it can't be done by hand, the problem can still be
resolved via the computer programs (such as MAPLE or Mathematica),
powerful enough to handle such type of calculations.

The second question is the necessity of the negative sign in front
of the corresponding terms with $\lambda$ for the case of
$|\phi|^4$ model (Sec. 2 and 3). Is it possible to construct exact
solution with positive sign (i.e.with positive coupling)? The
answer is yes, but in this case we'll have to deal with the
singular solution. This singularity is located along the hyperbola
and the physical meaning of such solution is unclear for us.

The last question is connected with the amazing fact that the dimension $D=4$ stands out.
Whether there is a simple {\em explanation} of this fact?

We conclude that  one can use the generalized Fubini instantons in
quantum cosmology only for the case of the  single infinite extra
dimension.


\end{document}